\documentclass[english]{article}
\usepackage[T1]{fontenc}
\usepackage[latin1]{inputenc}
\usepackage{babel}
\usepackage{amsmath}
\usepackage{amssymb}
\usepackage{amsfonts}
\usepackage{amsthm}
\usepackage[all,cmtip]{xy}
\usepackage{graphicx}
\usepackage{mathrsfs}

\theoremstyle{definition} 

\makeatletter
\@addtoreset{equation}{section}

\makeatother

\begin{document}

\title{Homotopy and Path Integrals}
\author{Fumika Suzuki\footnote{ \emph{E-mail address:}  \texttt{fumika@physics.ubc.ca}}\\\\ \textit{Department of Physics and Astronomy, University of British Columbia}}
\maketitle

\textbf{Abstract} This is an introductory review of the connection between homotopy theory and path integrals, mainly focus on works done by Schulman [23] that he compared path integral on $SO(3)$ and its universal covering space $SU(2)$, DeWitt and Laidlaw [15] that they proved the theorem to the case of path integrals on the multiply-connected topological spaces. Also, we discuss the application of the theorem in Aharonov-Bohm effect given by [20,24]. An informal introduction to homotopy theory is provided for readers who are not familiar with the theory.\\


\section{Introduction}
\quad Homotopy theory is the branch of algebraic topology and its main tools to study the properties of topological spaces are paths and loops. On the other hand, path integral is a technique in quantum mechanics to calculate the transition amplitude of a physical system from one point to another by summing over all paths connecting two points. It was suggested that there are interesting relations between two subjects by Schulman [23], DeWitt and Laidlaw [15].

In this paper, firstly we will review some basic homotopy theory in an informal way for readers who are not familiar with it (section 2). Then, we will explain the theorem which was proved by DeWitt and Laidlaw [15] which describes what happens to path integral if there are multiple homotopy classes of paths from one point to another. The application of the theorem to the statistics of identical particles given by them will be also discussed (section 3). Then, an example found by Schulman [23] will be explained, which indicates a connection between path integrals and the topological structure of spaces whose properties are described using homotopy theory  by comparing the topological structure of $SU(2)$ and $SO(3)$ and calculating the propagators on those spaces. (section 4) Finally, we will discuss the application of the theorem by DeWitt and Laidlaw in Aharonov-Bohm effect which was studied by Morandi, Menossi and Schulman. [20,24] (section 5)

\section{An Informal Introduction to Homotopy Theory}

\quad In this section, we introduce some basic concepts of homotopy theory which will be used in the following sections. Since many important theorems, however not used in this review are eliminated, it is recommended to refer to some textbooks of algebraic topology such as [13,14,19] for further understanding. 

Homotopy theory is the branch of algebraic topology and we will deal with properties of \emph{topological spaces}. Topological space is a generalization of Euclidean spaces in which we use set theory rather than the concept of distance to describe ideas such as closeness or limits.\\

\textbf{Definition 2.1 (Topology and Topological spaces):} \emph{A topology on a set $X$ is a collection $\mathcal{T}$ of subsets of $X$ having the following properties:}\\

($T_{1}$) $\emptyset$ and $X$ are elements of $\mathcal{T}$.

($T_{2}$) The union of any collection of elements in $\mathcal{T}$ is in $\mathcal{T}$.

($T_{3}$) The intersection of any finite collection of elements in $\mathcal{T}$ is in $\mathcal{T}$.\\

Then a \emph{topological space} is an ordered pair $(X, \mathcal{T})$ consisting of a set $X$ and a topology $\mathcal{T}$.\\

Two topological spaces are topologically identical if there exists a continuous deformation from one to another. One of the famous examples is that a topologist can't distinguish a coffee mug from a doughnut since we can form one into another if it is made of modeling clay. 
The continuous deformation such as stretching or bending is called homeomorphism and mathematically defined as follows:\\

\textbf{Definition 2.2 (Homeomorphism):} \emph{Two topological spaces $X$ and $Y$ are said to be homeomorphic (topologically equivalent) if there exists bijection $f: X \rightarrow Y$ which is continuous and has continuous inverse $f^{-1}: Y \rightarrow X$
. $f$ is called homeomorphism.}\\

Roughly speaking, topological equivalence can only be destroyed by tearing or gluing parts. Now, let us see what happens if we glue some parts of topological spaces.\\

\textbf{Definition 2.3 (Quotient spaces):} \emph{Let $X$ be a topological space and let $\sim$ be an equivalence relation on $X$. Define the equivalence class of $x \in X$ by}\\

\begin{center}
$[x] = \{ y \in X : y \sim x \}$
\end{center}

Then the \emph{quotient space} $X/ \sim$ is defined as the set of equivalence classes of the relation $\sim$ :

\begin{center}
$X/ \sim = \{[x] : x \in X \}$
\end{center}

Those readers who are not familiar with group theory can think about the quotient space $X / \sim$ as a new space which is created from $X$ by gluing $x$ to any $y$ in $X$ that satisfies $y \sim x$. Let us show you some examples:\\

\textbf{Example 2.3:} Let $X$ be a square $[-1,1] \times [-1,1]$

\begin{center}
$X = \{ (x,y) | -1 \leq x \leq 1 , -1 \leq y \leq 1 \}$
\end{center}

(i) Define the equivalence classes $\sim$ by $(1, t) \sim (-1,t)$ for all $t \in [-1,1]$. Then $X/ \sim$ is a cylinder.

\begin{figure}[htbp]
 \begin{center}
  \includegraphics[width=50mm]{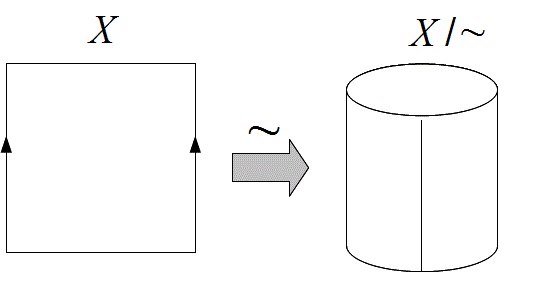}
 \end{center}
 \caption{Cylinder}
 \label{fig:one}
\end{figure}

(ii) Define the equivalence classes $\sim$ by $(1,t) \sim (-1,-t)$ for all $t \in [-1,1]$. Then $X/ \sim$ is M\"{o}bius band. 

\begin{figure}[htbp]
 \begin{center}
  \includegraphics[width=60mm]{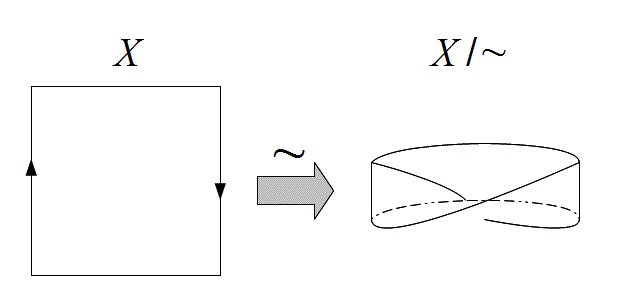}
 \end{center}
 \caption{M\"{o}bius band}
 \label{fig:two}
\end{figure}

(iii) Define the equivalence classes $\sim$ by $(1,t) \sim (-1,t)$ for all $t \in [-1,1]$ and $(s,1) \sim (s,-1)$ for all $s \in [-1,1]$. Then $X/ \sim$ is Torus.\\

\begin{figure}[htbp]
 \begin{center}
  \includegraphics[width=70mm]{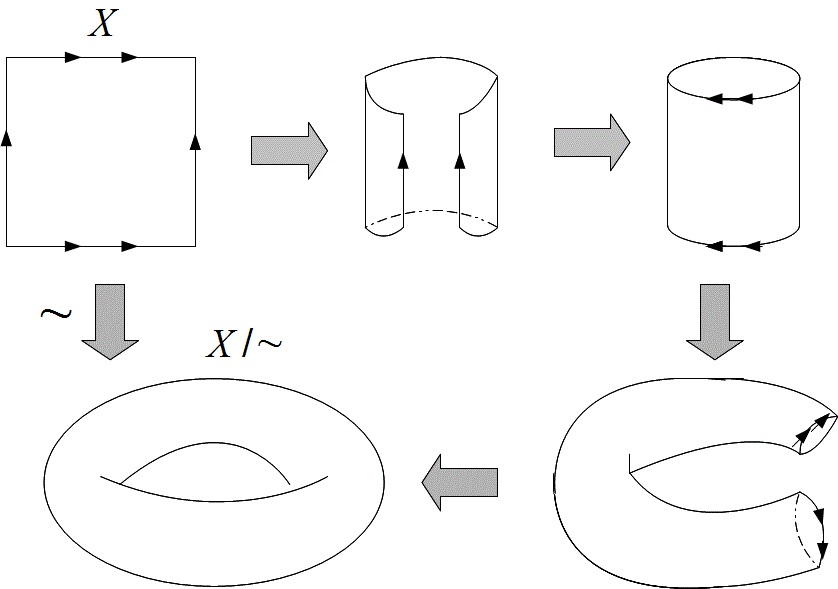}
 \end{center}
 \caption{Torus}
 \label{fig:3}
\end{figure}

\newpage
Now we introduce \emph{paths} which is a central tool to study the properties of topological spaces in homotopy theory.\\

\textbf{Definition 2.4 (Path):} \emph{Let $X$ be a topological space and let $x,y \in X$. Then a path in $X$ from $x$ to $y$ is a continuous function $\alpha : I \rightarrow X$ where $I=[0,1]$ with $\alpha (0)=x$ and $\alpha (1)=y$.}\\

\textbf{Example 2.4:}  

(i) $\alpha : I \rightarrow \mathbb{R}^2$, $\alpha(s)=(\cos \pi s, \sin \pi s)$ ($s \in I$) is a path in $\mathbb{R}^2$ from $(1,0)$ to $(-1,0)$\\
\begin{figure}[htbp]
 \begin{center}
  \includegraphics[width=50mm]{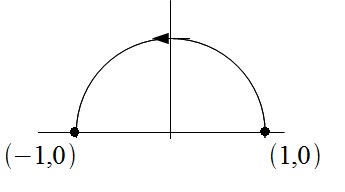}
 \end{center}
 \caption{Path (i)}
 \label{fig:}
\end{figure}

(ii) $\alpha : I \rightarrow \mathbb{R}^2$ $\alpha (s) = (\cos 2 \pi s , \sin 2 \pi s)$ $(s \in I$) is a path from $(1,0)$ to $(1,0)$ which is called "loop".\\

(iii) $\alpha : I \rightarrow X$ $\alpha(s)=x_0$ $(x_0 \in X, s \in I)$ is a constant path (or a constant loop at $x_0$).\\

\newpage

\textbf{Definition 2.5 (Path-connected):} \emph{$X$ is path-connected if there is a path in $X$ from $x$ to $y$ for all $x, y \in X$. A path-connected component of $X$ is an equivalence class under the equivalence relation $x \sim y$.}\\

\textbf{Theorem 2.6:} \emph{If $X$ is path-connected and $f: X \rightarrow Y$ is continuous then $f(X)$ is path-connected. If $f$ is surjective then $Y$ is path-connected.}\\

\emph{Proof:} Let $y_1 = f(x_1), y_2=f(x_2)$ $(x_1,x_2 \in X)$. Then there exists a path from $x_1$ to $x_2$, $\alpha: I \rightarrow X$ with $\alpha(0)=x_1, \alpha(1)=x_2$.

Then $f \circ \alpha : I \rightarrow Y$ is a path with $(f \circ \alpha) (0) =y_1 , (f \circ \alpha)(1)=y_2$

Since $f \circ \alpha$ is a composition of two continuous maps, it is continuous.
\begin{flushright}$\Box$\end{flushright}
\begin{figure}[htbp]
 \begin{center}
  \includegraphics[width=80mm]{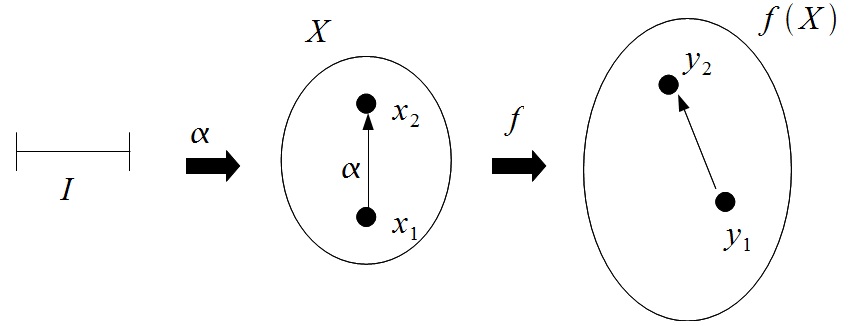}
 \end{center}
 \caption{Theorem 2.6}
 \label{fig:}
\end{figure}

\textbf{Corollary 2.7:} \emph{If $X$ is homeomorphic to $Y$ then}\\

(i) $X$ is path-connected if and only if $Y$ is.

(ii) The number of path-connected components of $X,Y$ are equal.\\

\textbf{Example 2.7:} $\mathbb{R}$ is not homeomorphic to $\mathbb{R}^2$

Suppose there exists homeomorphism $f : \mathbb{R} \rightarrow \mathbb{R}^2$

Then $f : \mathbb{R} \setminus \{0\} \rightarrow \mathbb{R}^2 \setminus \{f(0)\}$ is also a homeomorphism.

However, $\mathbb{R} \setminus \{0\}$ is not path-connected. $\mathbb{R}^2 \setminus \{0\}$ is path-connected.

It is contradiction. There exists no homeomorphism $f: \mathbb{R} \rightarrow \mathbb{R}^2$.

\begin{figure}[htbp]
 \begin{center}
  \includegraphics[width=65mm]{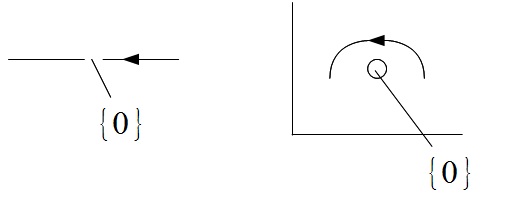}
 \end{center}
 \caption{Not path-connected (left), path-connected (right)}
 \label{fig:}
\end{figure}

\newpage
\textbf{Definition 2.8 (Homotopy of paths):} \emph{Let $\alpha : I \rightarrow X, \beta: I \rightarrow X$ be paths in $X$ from $x$ to $y$ then $\alpha$ is homotopic to $\beta$ if there is a continuous function $H: I \times I \rightarrow X$ such that}

\begin{center}
$H(s,0) = \alpha (s) \quad (s \in I)$, $H(s,1)=\beta (s) \quad (s \in I)$

$H(0,t)=x \quad (t \in I)$, $H(1,t)=y \quad (t \in I)$
\end{center}

Suppose $\alpha _{t} (s) = H(s,t)$. $\alpha _0 = \alpha$, $\alpha_1 = \beta$. Then $\alpha_{t}$ is 1-parameter family of paths deforming $\alpha$ to $\beta$ as $t$ gets from $0$ to $1$.

$H$ is called \emph{homotopy} from $\alpha$ to $\beta$. We write $\alpha \sim \beta$ for $\alpha$ is homotopic to $\beta$.

\begin{figure}[htbp]
 \begin{center}
  \includegraphics[width=70mm]{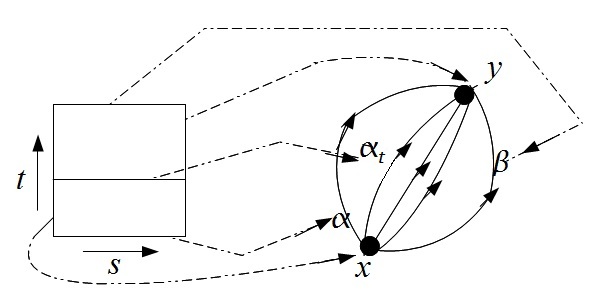}
 \end{center}
 \caption{Homotopy of paths}
 \label{fig:}
\end{figure}

\textbf{Example 2.8:} 

(i) Let $\alpha$ and $\beta$ are paths in a disk $D^2$ such that

\begin{center}
$\alpha : I \rightarrow D^2 \quad \alpha (s) = (\cos \pi s, \sin \pi s)$

\quad $\beta : I \rightarrow D^2 \quad \beta (s) = (\cos \pi s , - \sin \pi s)$
\end{center}

Define $H: I \times I \rightarrow D^2$ by $H(s,t) = (\cos \pi s, (1-2t) \sin \pi s)$

Since $\cos ^2 \pi s + (1-2t)^2 \sin ^2 \pi s \leq \cos ^2 \pi s + \sin ^2 \pi s =1$, 

$H(s,t)$ is in $D^2$ for all $(s,t) \in I \times I$.

We have $H(s,0) = \alpha (s) , H(s,1) = \beta (s), H (0,t)=(1,0) , H(1,t)=(-1,0)$.

Thus $H$ is a homotopy from $\alpha$ to $\beta$, so $\alpha \sim \beta$.

\begin{figure}[htbp]
 \begin{center}
  \includegraphics[width=50mm]{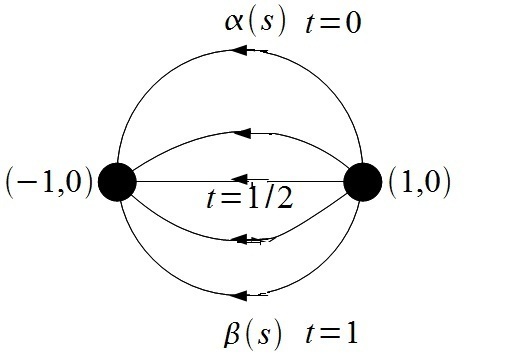}
 \end{center}
 \caption{Homotopy}
 \label{fig:}
\end{figure}

\newpage
Note that if we change a disk $D^2$ into an annulus by making a hole, then any attempts to find $H$ will fail and $\alpha$ and $\beta$ are not homotopic on an annulus.\\

Another central tool in homotopy theory is \emph{loop}.\\

\textbf{Definition 2.9 (Loop)} \emph{A loop (based) at $x$ is a path in $X$ from $x$ to $x$ which is a continuous function $\alpha : I \rightarrow X, \alpha (0)= \alpha (1) =x$}\\

If $\alpha , \beta , \gamma , \delta$ are all loops at $x$, we have

(1) $(\alpha \ast \beta) \ast \gamma \sim \alpha \ast (\beta \ast \gamma)$

(2) $\alpha \sim \gamma , \beta \sim \delta \rightarrow \alpha \ast \beta \sim \gamma \ast \delta$

(3) $ e_{x} \ast \alpha \sim \alpha \sim \alpha \ast e_{x}$

(4) $ \alpha \sim \beta \rightarrow \bar{\alpha} \sim \bar{\beta}$

(5) $\alpha \ast \bar{\alpha} \sim e_{x} \sim \bar{\alpha} \ast \alpha$\\

\textbf{Definition 2.10 (Fundamental group)} \emph{The fundamental group of a topological space $X$ with base point $x$ is}

\begin{center}
$\pi_1 (X,x) =\{$all loops $\alpha : I \rightarrow X$ where $\alpha$ is based at $x \}$

i.e., The elements of $\pi_1(X,x)$ are the homotopy classes of loops at $x$.

\end{center}

\textbf{Theorem 2.11} \emph{$\pi_1 (X,x)$ is a group.}\\

\emph{Proof:} 

Let us denote the equivalence class (homotopy class) of loops at $x$ which are homotopic to $\alpha$ by $[\alpha]$. Then $[\alpha]=[\beta]$ means $\alpha \sim \beta$.

Then the multiplication of the fundamental group is defined by $[\alpha][\beta]=[\alpha * \beta]$ which is well defined. 

i.e., $[\alpha]=[\gamma], [\beta]=[\delta]$ implies $[\alpha * \beta]=[\gamma * \delta]$.\\

If $[\alpha], [\beta]$ are the homotopy class of loops at $x$, then $[\alpha * \beta]$ is also the homotopy class of loops at $x$.

The identity is $[e_{x}]$ since $[e_{x}][\alpha]=[\alpha]=[\alpha][e_{x}]$ by (3).

The inverse of $[\alpha]$ is $[\bar{\alpha}]$ since $[\alpha]=[\beta]$ implies $[\bar{\alpha}] =[\bar{\beta}]$ by (4) which tells the inverse is well-defined. From (5), we have

\begin{center}
$[\alpha][\bar{\alpha}]=[\alpha * \bar{\alpha}] =[e_{x}]$ and $[\bar{\alpha}][\alpha]=[e_{x}]$
\end{center}

Associativity follows from (1).

$([\alpha][\beta])[\gamma]=[\alpha * \beta][\gamma]=[(\alpha * \beta) * \gamma]=[\alpha *  (\beta * \gamma)]=[\alpha][\beta * \gamma]=[\alpha]([\beta][\gamma])$

\begin{flushright}$\Box$\end{flushright}

\newpage

\textbf{Example 2.10:} 

(i) For any $x \in \mathbb{R}^{n}, \pi_1 (\mathbb{R}^2 ,x)$ is the trivial group since if $\alpha$ is any loop then $\alpha \sim e_{x}$.\\

(ii) Similarly for any $n$-dimensional ball $D^{n} , \pi_1 (D^{n} , x)$ is trivial.

Those path-connected spaces with a trivial fundamental group is called \emph{simply-connected}.\\

(iii) For any $x \in S^1$ (circle), $\pi_1 (S^1,x) \simeq \mathbb{Z}$.

Each homotopy class consists of all loops $\alpha _{n}$ which wind around the circle $n$ times, $\alpha_{n} = e^{2 \pi i ns}$ $(n \in \mathbb{Z},s \in [0,1])$. i.e., any other loop is homotopic to $\alpha_{n}$ for some $n$.

Since the product of a loop which winds around $m$ times and another that winds around $n$ times is a loop which winds around $m + n$ times, the fundamental group is isomorphic to the additive group of integers $\pi_1 (S^1,x) \simeq \mathbb{Z}$.

Those spaces that are connected but not simply-connected are called \emph{multiply-connected}.\\

(iv) For any $x \in S^{n} , \pi_1 (S^{n} ,x)$ $(n-$sphere, $n >1)$ is the trivial group since we can continuously deform any loops on $n$-sphere ($n >1$) into a point.\\

\begin{figure}[htbp]
 \begin{center}
  \includegraphics[width=40mm]{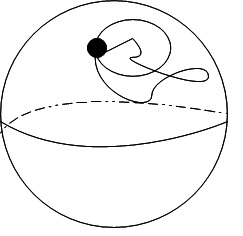}
 \end{center}
 \caption{Loop on sphere}
 \label{fig:}
\end{figure}

\newpage

\section{Path Integrals in Multiply-connected Spaces}

\quad Paths and loops which appear in homotopy theory remind physicists about path integral. Let $a$ and $b$ be some points in the \emph{configuration space} $X$ of some physical system. In this review, by configuration space, we mean that it is the space of possible positions of the whole system and should not be confused with the \emph{phase space}. For example, the configuration space of the physical system of $n$ free particles is $\mathbb{R}^{3n}$.  Then path integral is a way of calculating the transition amplitude of a physical system from some point $a$ to $b$ in the configuration space $X$ by summing over all possible paths from $a$ to $b$ in $X$. (For good introduction to path integral, reader can refer to [8,17].) However, path integral is defined only for the paths in the same homotopy class in the configuration space. Therefore, although we do not have any problem when the configuration space is simply-connected space such as $\mathbb{R}^2$ since we have only one homotopy class of paths from $a$ to $b$ denoted $[q (a,b)]$, the problem arises when the configuration space has a hole in it and multiply-connected such as $\mathbb{R}^2 \setminus \{0\}$ ($\mathbb{R}^2$ where a point $\{0\}$ is removed.) and there are multiple homotopy classes of paths such that $[q_1 (a,b)]$, $[q_2 (a,b)]$ or $[q'_{n} (a,b)]$. (i.e., a path which goes from $a$ to $b$ after winding around a hole $n$ times where $n \in \mathbb{Z}$ with $n \not=0$ whose sign indicates the winding direction.) (Figure 10)

To calculate the transition amplitude in such a configuration space, we need to sum over the contributions from all such homotopy classes of paths. The theorem for path integral on these multiply-connected space was stated by DeWitt and Laidlaw [15].

\begin{figure}[htbp]
 \begin{center}
  \includegraphics[width=100mm]{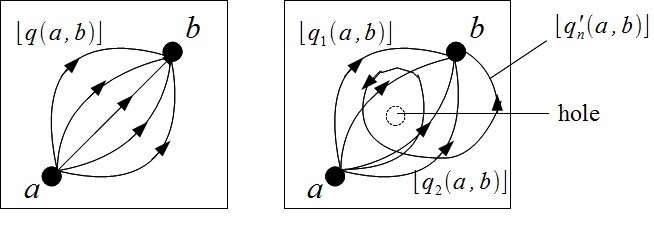}
 \end{center}
 \caption{Only one homotopy class of paths (left), multiple homotopy classes of paths (right)}
 \label{fig:}
\end{figure}

\textbf{Theorem 3.1 (The homotopy theorem for path integral):} \emph{Let the configuration space $X$ of a physical system be the topological space. Then the probability amplitude $K$ for a given transition is, up to a phase factor, a linear combination of partial probability amplitudes $K^{\alpha}$ obtained by integrating over paths in the same homotopy class in $X$:}[6]

\begin{center}
$K=\displaystyle\sum_{\alpha \in \pi_1 (X,x)} \chi (\alpha) K^{\alpha}$
\end{center}

where the coefficients $\chi (\alpha)$ form a one-dimensional unitary representation (or the character of a representation) of the fundamental group $\alpha \in \pi_1 (X,x)$.

A complete proof of the theorem is given in [15] and [11] provides some simple explanation of the proof. Here, we briefly explain the proof which is given in those references.\\

\emph{Proof:} Since we can not include paths of different homotopy classes in path integral, (i.e., path integral is defined only for the paths in the same homotopy class), we "assume" that we can include all paths by taking the sum of the different amplitudes $K^{\alpha}$ for each homotopy class with some weight factors $\chi (\alpha)$:

\begin{center}
$K= \displaystyle\sum _{\alpha \in \pi_1 (X,x)} \chi (\alpha) K^{\alpha}$
\end{center}

Then the weight factors $\chi (\alpha)$ form a one-dimensional unitary representation of the fundamental group. It was proved as follows.

Let $a, b$ be any two points in the configuration space $X$ and let $[q(a,b)]$ be the homotopy classes of paths from $a$ to $b$ which are homotopic to $q(a,b)$.

Let the set of all such homotopy classes be $\pi [X,a,b]$. (i.e., $\pi [X,a,b]$ includes all different homotopy classes of paths from $a$ to $b$.)

Let $x$ be some fixed point in $X$, and let $\pi_1 (X,x)$ be the set of homotopy classes of loops based at $x$.

Then we can construct the mapping $f_{ab}$ from $\pi_1 (X,x)$ to $\pi (X,a,b)$ for every $a, b \in X$ such that

\begin{center}
$f_{ab} : \pi_1 (X,x) \rightarrow \pi (X,a,b)$
\end{center}

by

\begin{center}
$f_{ab} (\alpha) =[C^{-1} (a)] \alpha [C(b)]$
\end{center}

where $\alpha$ is one of the loops based at $x$, $\alpha \in \pi_1 (X,x)$ and $C(a)$ denotes an arbitrarily chosen path from $x$ to $a$ for every $a \in X$.

\begin{figure}[htbp]
 \begin{center}
  \includegraphics[width=50mm]{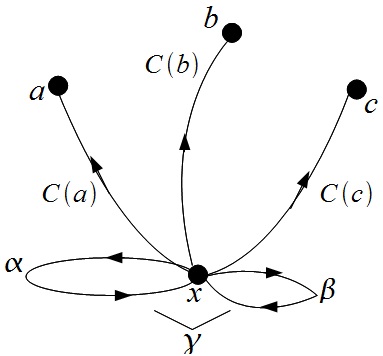}
 \end{center}
 \caption{Theorem 3.1 (i)}
 \label{fig:}
\end{figure}

Now, let $\alpha,\beta, \gamma \in \pi_1(X,x)$ be the loops based at $x$ and $\alpha * \beta = \gamma$. Therefore $\gamma$ is the loop such that it goes around the loop $\alpha$ first and then the loop $\beta$ and comes back to $x$.

Let $a,b,c$ are points in $X$. Then we can describe the path from $a$ to $c$ using $\gamma$ by

\begin{center}
$f_{ac} (\gamma) = [C^{-1} (a)] \gamma [C(c)]$
\end{center}

However, every path $q \in f_{ac} (\gamma)$ can be split into two paths $q_1 \in f_{ab} (\alpha)$ and $q_2 \in f_{bc} (\beta)$ since

\begin{center}
$f_{ac} (\gamma) = [C^{-1} (a)] \gamma [C(c)]$

$=[C^{-1} (a)]\alpha [C (b)] [C^{-1} (b)] \beta [C (c)]$

\end{center}

What it says is that the path that goes from $a$ to $x$, goes around $\alpha$ and $\beta$ (which is the loop $\gamma$) and then arrives at $c$ can be split into the path that goes from $a$ to $x$, goes around $\alpha$ and then goes to $b$  and goes around $\beta$ after coming back to $x$ and arrives at $c$.

Now for $K(c, t_{c} ; a , t_{a})$, we can combine amplitudes for occuring in succession time:

\begin{center}
$K(c,t_{c} ; a, t_{a}) =\int _{x_{b}} K(c, t_{c} ; b, t_{b}) K(b, t_{b} ; a, t_{a}) dx_{b}$ if $t_{a} < t_{b} < t_{c}$
\end{center}

This rule can be derived from the property of the action $S[c,a]=S[c,b]+S[b,a]$.

Then by the assumption, we have

\begin{center}
$\displaystyle\sum _{\gamma \in \pi_1 (X,x)} \chi (\gamma) K^{\gamma}(c,t_{c} ; a, t_{a}) $

$= \displaystyle\sum _{\alpha , \beta \in \pi_1 (X,x)} \chi (\alpha) \chi (\beta) \int  K^{\beta} (c,t_{c} ; b,t_{b}) K^{\alpha} (b,t_{b} ; a,t_{a})dx_{b}$
\end{center}

Since

\begin{center}
$K^{\gamma}(c,t_{c} ; a, t_{a})  = \int  K^{\beta} (c,t_{c} ; b,t_{b}) K^{\alpha} (b,t_{b} ; a,t_{a})dx_{b}$
\end{center}

we have $\chi (\gamma) = \chi (\alpha * \beta) = \chi (\alpha) \chi (\beta)$.\\

\begin{figure}[htbp]
 \begin{center}
  \includegraphics[width=60mm]{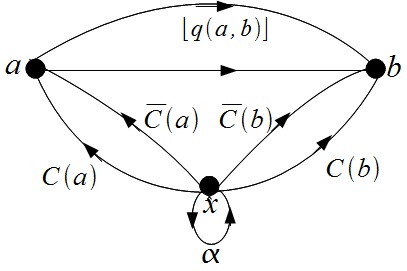}
 \end{center}
 \caption{Theorem 3.1 (ii)}
 \label{fig:}
\end{figure}

Now, let $\bar{C} (a)$ be an arbitrary chosen path from $x$ to $a$ which is different from $C (a)$.

Then we have a map $\bar{f}_{ab}$ such that

\begin{center}

$\bar{f}_{ab} (\alpha) = [\bar{C} ^{-1} (a)] \alpha [\bar{C} (b)]$

$=[C^{-1} (a)] [C(a)] [\bar{C} ^{-1} (a)] \alpha [\bar{C} (b)] [C^{-1} (b)] [C(b)]$

$=[C^{-1} (a)] \lambda \alpha \mu [C(b)] =f_{ab} (\lambda \alpha \mu)$
\end{center}

where $\lambda$ and $\mu$ are the loops based at $x$. (i.e., $\lambda =[C(a) \bar{C} ^{-1} (a) ], \mu= [\bar{C} (b) C^{-1} (b)] \in \pi_1 (X,x)$).

We can see that the mapping $f_{ab}$ labels each homotopy class paths from $a$ to $b$ with an element of the fundamental group and the above is the transformation from the labelling $f_{ab}$ to another labelling $\bar{f} _{ab}$.

Since the absolute value of the total amplitude $K$ is invariant under this transformation or the choice of labelling, we have

\begin{center}
$|K(b,t_{b} ; a, t_{a})| =| \displaystyle\sum _{\alpha \in \pi_1(X,x)} \chi (\alpha) K^{\alpha} (b,t_{b} ; a, t_{a})| $

$= |\displaystyle\sum_{\lambda \alpha \mu \in \pi_1 (X,x)} \chi (\lambda \alpha \mu) K^{\alpha} (b,t_{b} ; a, t_{a})|=|\displaystyle\sum _{\lambda \alpha \mu \in \pi_1(X,x)} \chi (\alpha) \chi (\lambda \mu) K^{\alpha} (b,t_{b} ; a, t_{a})|$
\end{center}

Then if $\chi$ satisfies the following properties the transition amplitude is unchanged:

\begin{center}
$\chi (\alpha) \chi (\beta) = \chi (\alpha \beta)$ with $|\chi (\alpha)| = 1$ for any $\alpha, \beta \in \pi_1 (X,x)$
\end{center}

This implies that the weight factors $\chi$ form a one-dimesnional unitary representation of the fundamental group.

\begin{flushright}$\Box$\end{flushright}

One application of this theorem was also discussed by DeWitt and Laidlaw [15].\\

\newpage
\textbf{Application 3.1:} Let us consider the physical system with $n$ free indistinguishable spinless particles in $d$-dimensional space $\mathbb{R}^{d}$. Then a point of the configuration space $X$ of such a sytem is the set

\begin{center}
$x=\{\textbf{x}_1, \ldots, \textbf{x}_{n}\} \in X$ with $\textbf{x}_{i} \in \mathbb{R}^{d}$ 
\end{center}

where $\textbf{x}_{i} \not= \textbf{x}_{j}$ if $i \not= j$ since no two particles can occupy the same position (i.e., particles are assumed to be spinless.) and the set $x=\{\textbf{x}_1,\ldots,\textbf{x}_{n}\}$ is unordered (i.e., $x=\{\textbf{x}_1, \textbf{x}_2, \ldots, \textbf{x}_{n}\}=\{\textbf{x}_2,\textbf{x}_1, \ldots, \textbf{x}_{n}\}$) since particles are indistinguishable.

To find the fundamental group of this configuration space $X$, we need to make a loop in $X$. Let $x^0=\{\textbf{x}_1^0, \ldots, \textbf{x}_{n}^0\}$ be the base point of a loop $\alpha$. Then $\alpha \in \pi_1 (X, x_0)$ is defined by [3]

\begin{center}
$\alpha (t)=\{\textbf{x}_1^0, \textbf{x}_2^0, \ldots , \textbf{x}_{i}^0(t), \textbf{x}_{i+1}^0, \ldots , \textbf{x}_{j-1}^0, \textbf{x}_{j}(t), \textbf{x}_{j+1}^0, \ldots, \textbf{x}_{n}^0\}$

with $\textbf{x}_{i}(0)=\textbf{x}_{i}^0, \textbf{x}_{i}(1)=\textbf{x}_{j}^0, \textbf{x}_{j}(0)=\textbf{x}_{j}^0, \textbf{x}_{j} (1) =\textbf{x}_{i}^0$ for $0 \leq t \leq 1$.
\end{center}

Therefore, $\alpha$ interchanges a particle $i$ and $j$ using time $t$ and it is a loop since the set $x$ is unordered. Then it is discussed in [3] that the fundamental group $\pi_1 (X, x_0) \simeq S_{n}$ for $d \geq 3$ as follows. Since the loop $\alpha$ interchanges two particles $i$ and $j$, it is identified with the transpositions $s_{ij}$. (i.e., a function that swaps two elements of a set.) Let $s_{i,i+1}=\sigma_{i} (1 \leq i \leq n-1)$, then for $d \geq 3$, we have\\

(i) $\sigma_{i}\sigma_{i+1}\sigma_{i}=\sigma_{i+1}\sigma_{i}\sigma_{i+1}$

(ii) $\sigma_{i}\sigma_{j}=\sigma_{j}\sigma_{i}$ if $|i-j| \geq 2$

(iii) $\sigma_{i}^2=e$\\

For (i), let $i=1$ then what it says is that the operation which exchanges particles $1$ and $2$, $2$ and $3$, then $1$ and $2$ again is same as the operation which exchanges particles $2$ and $3$ , $1$ and $2$, then $2$ and $3$. As we can see in the Figure 13, the loops associated $\sigma_1 \sigma_2 \sigma_3$ and $\sigma_2 \sigma_1 \sigma_2$ are homotopic. Notice that the Figure 13 as well as Figure 14 describes the spatial dimensions higher than two since particles collide with each other before the exchange in one-dimensional space. (ii) is showed in the similar way and it just says that the operation which interchanges particles $i$ and $i+1$ after interchanging $j$ and $j+1$ is same as the operation which interchanges $i$ and $i+1$ before interchanging $j$ and $j+1$.

\begin{figure}[htbp]
 \begin{center}
  \includegraphics[width=100mm]{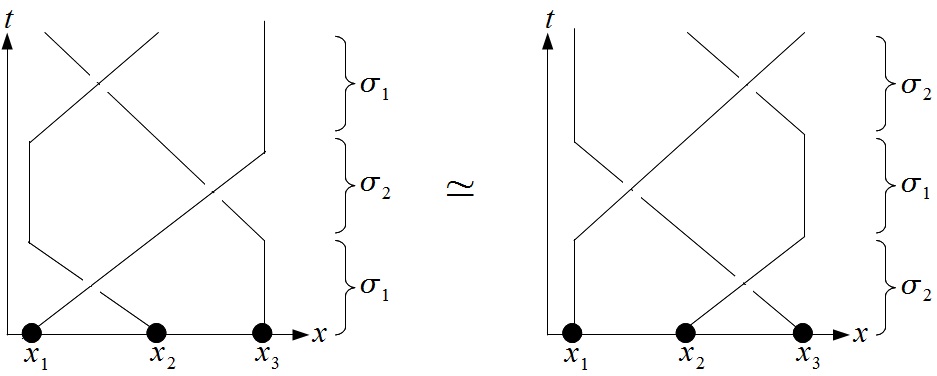}
 \end{center}
 \caption{(i) for $i=1$ and $n=3$.}
 \label{fig:}
\end{figure}

Interesting fact arises when we consider the property (iii). (iii) indicates that the operation which interchanges two particles $i$ and $i+1$ twice is same as doing nothing. Now, the operation which interchanges the position of two particles twice is topologically equivalent to the operation which one particle looping around the other. (Figure 14) In three or higher spatial dimensions, (i.e., $d \geq 3$) it is possible for this loop to be shrunk to a point by escaping to a higher dimension from a two-dimensional plane. However, in two-dimensional space, the loop can not be shrunk to a point since there exists $x_2$ as a hole which prevent it. (Figure 15)

\begin{figure}[htbp]
 \begin{center}
  \includegraphics[width=150mm]{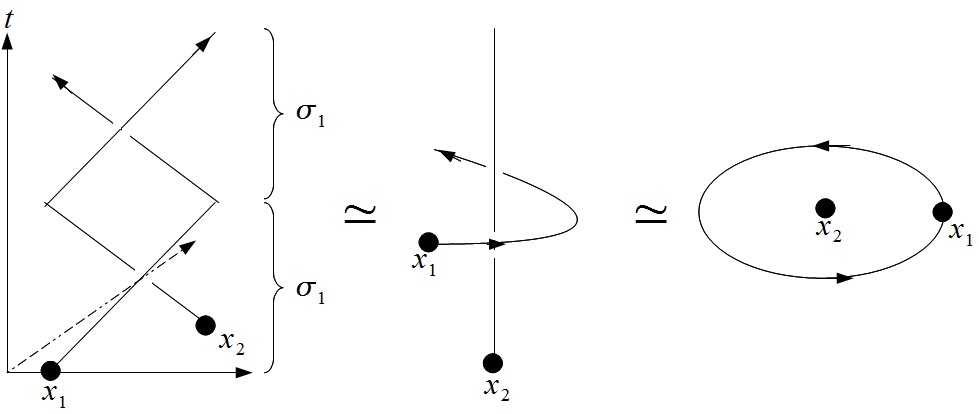}
 \end{center}
 \caption{(iii) for $i=1$ and $n=2$.}
 \label{fig:}
\end{figure}

\begin{figure}[htbp]
 \begin{center}
  \includegraphics[width=100mm]{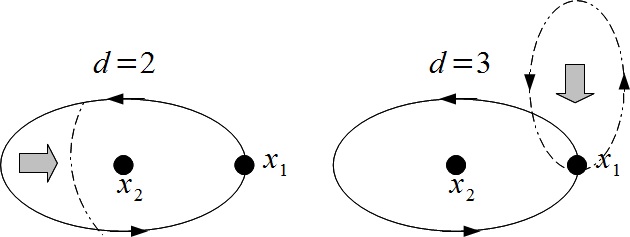}
 \end{center}
 \caption{(iii) Difference between two-dimensional and three-dimensional space.}
 \label{fig:}
\end{figure}

Therefore, although the properties (i), (ii) and (iii) hold in $d$-dimensional space where $d \geq 3$, in two-dimensional space, only the properties (i) and (ii) hold. The group with generators satisfying (i), (ii) and (iii) is known to be the symmetric group $S_{n}$. Therefore, the fundamental group $\pi_1 (X,x_0)$ is isomorphic to $ S_{n}$ in three or higher spatial dimensions. On the other hand, the group satisfying only (i) and (ii) is called the braid group $B_{n}$ and the fundamental group of two spatial dimensions is isomorphic to $B_{n}$. Since the property (iii) fails in two-dimensional space, identical particles can not be labelled as bosons or fermions in the space and can have any phase factors. Those particles are called \emph{anyons}.

Thus, if we have a physical system of $n$ "distinguishable" particles in $\mathbb{R}^{d}$, then the configuration space $\tilde{X}$ of such a system is
\begin{center}
$\tilde{X}(n,d)=\{x=(\textbf{x}_1, \ldots, \textbf{x}_{n}) ; \textbf{x}_{i} \in \mathbb{R}^{d}$ and $\textbf{x}_{i} \not= \textbf{x}_{j}$ if $i \not= j\}$
\end{center}

where $x=(\textbf{x}_1, \ldots, \textbf{x}_{n})$ is an ordered $n$-tuple. 

Then, $\tilde{X}(n,1)$ is not connected, $\tilde{X}(n,2)$ is multiply-connected and $\tilde{X}(n,d)$ with $d \geq 3$ is simply-connected. 

For a physical system of $n$ indistinguishable particles in $3$-dimensional space, since points which differ by their interchange or permutations belonging to $S_{n}$ are identified, the configuration space $X$ of such a system is the quotient space:

\begin{center}
$X=\tilde{X}(n,3) / S_{n}$
\end{center}

Mathematically, it is known that there are only two one-dimensional unitary representations of the symmetric group $S_{n}$ and therefore we have [6]\\

$K$(Bose)$=\displaystyle\sum_{\alpha \in \pi_1 (X,x_0)} \chi^{B} (\alpha) K^{\alpha}$ (symmetric propagator)

$K$(Fermi)$=\displaystyle\sum_{\alpha \in \pi_1(X,x_0)} \chi^{F} (\alpha) K^{\alpha}$ (antisymmetric propagator)

where $\chi^{B}$ and $\chi^{F}$ are two one-dimensional unitary representation of $S_{n}$:
\begin{center}
$\chi^{B}=+1$ for all permutations $\alpha \in S_{n}$
\end{center}

\[
\chi^{F}=
\begin{cases}
+1 &\text{for even permutations $\alpha \in S_{n}$}\\
-1 &\text{for odd permutations $\alpha \in S_{n}$}\\
\end{cases}
\]

Note that this approach to the statistics of indistinguishable particles has a connection with the study of the relation between topology and spin-statistics theorem. Some rigorous proof using relativity can be found in [7,25]. However, there are many attempts to prove this theorem without relativity. [3,4,10] Some discussion about the problem is given by Feynman. [9] Finkelstein and Rubenstein used the topological arguments to prove the theorem. [10] However, there exists some criticize such that these proofs require an additional assumption for quantum mechanics and it seems that creating a rigorous proof of the spin-statistic theorem in the nonrelativistic regime is still an open problem.\\

\newpage
\section{Path integral for a spinning particle}

\quad Another example which suggests the relation between homotopy and path integral was given by Schulman [23]. In his paper, he developed a path integral for a spinning particle.

Firstly, we need to know what is the configuration space for a spinning particle. Generally, spin is interpreted as a type of internal angular momentum. Therefore, according to Bopp and Haag [5], a spinning particle can be modelled as a charged rigid spherical ball with the internal dynamical variables represented by the Euler angles. Then the configuration space of such a rigid body in $\mathbb{R}^3$ is $\mathbb{R}^3 \times SO(3)$ where the position of the centre of mass is expressed by a vector $\mathbf{r} \in \mathbb{R}^3$ and the orientation of the body is represented by an orthogonal matrix $\mathcal{O} \in SO(3)$. $SO(3)$ is the 
group of all $3 \times 3$ orthogonal matrices such that elements are real and $\mathcal{O}^{T}\mathcal{O}=1, \det \mathcal{O}=1$ where $\mathcal{O}^{T}$ is the transpose of a matrix $\mathcal{O}$. A rotation in $\mathbb{R}^{n}$ is an element of $SO(n)$ which consists of $n \times n$ real matrices with $\mathcal{O}^{T}\mathcal{O}=1$ and $\det \mathcal{O}=1$ and it is sometimes called the \emph{rotation group}. 

It is known that an element of the group $SO(3)$ can be expressed in terms of a set of three parameters. (The reason is discussed below.) The Euler angles which describe the orientation of a rigid body is one example of such parameters. 

Now, we want to know the topological structure of the configuration space $\mathbb{R}^3 \times SO(3)$. We already know that $\mathbb{R}^3$ is simply-connected space and so there exists only one homotopy class of paths. An interesting discussion to determine the topological structure of $SO(3)$ can be found in [16] and [21] which is as follows.

A general $n \times n$ real matrix has $n^2$ entries and so is determined by $n^2$ real parameters. However, since $SO(n)$ has the orthogonality condition, if the elements of upper triangle of the matrix are determined, then the elements of the lower triangle are also fixed. Thus it has $\frac{n(n+1)}{2}$ constraints and so $SO(n)$ can be specified by $n^2 - \frac{n(n+1)}{2} =\frac{n(n-1)}{2}$ parameters. Then $SO(3)$ can be expressed by three parameters and so it is a three-dimensional manifold.
Now, any rotation is defined by some axis $\mathbf{n}$ and a right-handed turning through an angle $\theta$. [21] Therefore, the rotation can be represented by a vector with length $\theta$ where $0 \leq \theta \leq \pi$. Then, the collection of all such vectors forms a solid closed ball of radius $\pi$ in $\mathbb{R}^3$ denoted $D^3$. [16] Since a rotation by $\pi$ about the axis $\mathbf{n}$ is identical to the rotation by $\pi$ about $-\mathbf{n}$, the opposite points of the boundary $S^3$ of $D^3$ must be identified. Therefore, $SO(3) \simeq D^3 / \sim$.

This space is multiply-connected since it has two disjoint classes of loops on it, I and II:\\

Class I loops: It intersects with the boundary $S^3$ and so, for example, it contains all diameters of $D^3 /  \sim$.

Class II loops: It contains all internal loops which can be deformed  into a single point and make the trivial loops. \\

There exists the connection between a continuous rotation of an object which takes the object back to its initial orientation in $\mathbb{R}^3$ and these two classes of loops.

Class I loops represent a rotation through $2\pi$, while class II loops represent a rotation through $4\pi$. Class I loop can not be continuously deformed into the trivial loop which describes no motion of the object, however class II can. This fact is illustrated in many ways, for example, in Dirac's scissors problem which is explained in [21]. Since the fundamental group has just two elements, $\pi_1 (SO(3),x)\simeq \mathbb{Z}_2$ which is the group of integers mod $2$.

Now, it is interesting to consider the \emph{universal covering space} of $SO(3)$.\\

\textbf{Definition 4.1 (Covering space):} \emph{Let $X$ and $\tilde{X}$ be topological spaces. Then $\tilde{X}$ is a covering space of $X$ if there exists a surjective continuous map $p: \tilde{X} \rightarrow X$ satisfying the following conditions:}\\

There is an open neighbourhood $U$ of $x$ for each $x \in X$ such that

(i) $p^{-1} (U)$ is a disjoint union of open sets $\tilde{X}_{j} \subset \tilde{X}$.

(ii) Each $\tilde{X}_{j}$ is mapped homeomorphically onto $U$ via $p$.\\

$p$ is called a \emph{covering map}, the $\tilde{X}_{j}$ are \emph{sheets} of the covering of $U$ and $p^{-1} (x)$ for each $x \in X$ is the \emph{fiber} of $p$ above $x$. If $\tilde{X}$ is simply-connected, it is called the \emph{universal covering space}.\\

Informally, $\tilde{X}$ is obtained by unwrapping the identifications on the space $X$ maximally. For example, if $X$ is a circle $S^1$, then the paths on $S^1$ are identified with modulo $2\pi$. If we unwrap this identification, then we have the real line $\mathbb{R}$. Therefore the universal covering space of $S^1$ is a real line $\mathbb{R}$.

Now, the two-dimensional disk $D^2$ is topologically equivalent to the northern hemisphere of 2-sphere $S^2$ which is an ordinary sphere we often see in three-dimensional Euclidean space. Similarly, $D^3$ is the northern hemisphere of 3-sphere $S^3$. Therefore $SO(3) \simeq D^3/ \sim$ is same as the northern hemisphere of $S^3$ with opposite equatorial points are identified or all of $S^3$ with antipodal points identified.

The unit 3-sphere centred on the origin is the set of $\mathbb{R}^4$ defined by

\begin{center}
$S^3 =\{ (x_1,x_2,x_3,x_4) \in \mathbb{R}^4 : x_1^2+x_2^2+x_3^2+x_4^2=1\}$
\end{center}

and $SO(3)$ can be made by identifying antipodal points $(x_1,x_2,x_3,x_4) \sim (-x_1,-x_2,-x_3,-x_4)$.

The group which is topologically equivalent to $S^3$ is known as $SU(2)$. We can show this in the following way. $SU(2)$ is the group of all $2 \times 2$ unitary matrices with determinant $1$ and its elements are complex number. Let $U$ be a matrix $U \in SU(2)$ written by

\[
U =
\left( {\begin{array}{cc}
e&f \\
 g & h\\
 \end{array} } \right)
\]

with $U^{\dagger} U=1$ where $U^{\dagger}$ is the Hermitian adjoint of $U$ and $\det U=1$. Also, we have

\begin{center}
$U^{-1}=(\det U)^{-1} \begin{pmatrix}h&-f\\-g&e \end{pmatrix}$

\end{center}

Since $U^{-1} = U^{\dagger}$ by $U^{\dagger} U=1$, we have

\begin{center}
$\begin{pmatrix}h&-f\\-g&e \end{pmatrix} = \begin{pmatrix} \bar{e}&\bar{g}\\ \bar{f} & \bar{h} \end{pmatrix}$
\end{center}

where $\bar{e}$ is the complex conjugate of $e$. 

Then, $\bar{e}=h$ and $\bar{f}=-g$ and $U \in SU(2)$ takes the form:

\[
U (e,f)=
\left( {\begin{array}{cc}
e&f \\
 -\bar{f} & \bar{e}\\
 \end{array} } \right)
\]

with $|e|^2+|f|^2=1 , e,f \in \mathbb{C}$. (i.e., $\det U=1$ and so $\bar{e}e +\bar{f}f=1$)

Therefore $U \in SU(2)$ is represented by the vector $(e,f) \in \mathbb{C}^2$ of length 1.

Let $e=x_1 + x_2 i$ and $f=x_3 +x_4 i$, then the above condition implies

\begin{center}
$x_1 ^2+x_2^2 +x_3^2 +x_4^2=1$
\end{center}

which is the equation of the unit 3-sphere in $\mathbb{R}^4$ and $SU(2)$ is homeomorphic to the unit 3-sphere in $\mathbb{R}^4$. (i.e., there exists a continuous bijective map from $SU(2)$ to the unit 3-sphere with the continuous inverse.)

Also, we can discuss the fact by introducing \emph{quaternions}. Quaternions $\mathbb{H}$ is any number of the form $a \mathbf{1} +b \mathbf{i} + c\mathbf{j} + d \mathbf{k}$ where $a,b,c$ and $d$ are real numbers, $\mathbf{i}^2=\mathbf{j}^2=\mathbf{k}^2=\mathbf{ijk}=-1$.

Then $\mathbf{1},\mathbf{i},\mathbf{j}$ and $\mathbf{k}$ can be expressed by the following matrices:
\begin{center}
$\mathbf{1} = \begin{pmatrix} 1 & 0 \\ 0 & 1 \end{pmatrix} \quad \mathbf{i}= \begin{pmatrix} i & 0 \\ 0 & -i \end{pmatrix}$

$\mathbf{j}= \begin{pmatrix} 0 & 1 \\ -1 & 0 \end{pmatrix} \quad \mathbf{k}= \begin{pmatrix} 0 & i \\ i & 0 \end{pmatrix}$
\end{center}

Then clearly every matrix in $H \in \mathbb{H}$ is of the form

\[
H (x,y)=
\left( {\begin{array}{cc}
x&y \\
 -\bar{y} & \bar{x}\\
 \end{array} } \right)
\]

where $x=a+ib$ and $y=c+id$ and it is similar form to matrices $SU(2)$, however without any conditions, quaternions has dimension $4$, while $SU(2)$ has dimension $3$.

Since $\mathbb{R}^4$ can be regarded as the two-dimensional complex space $\mathbb{C}$ or the space of quaternions $\mathbb{H}$. We can rewrite the unit 3-sphere by

\begin{center}
$S^3 = \{ (z_1 , z_2) \in \mathbb{C}^2 : |z_1|^2 +|z_2|^2 =1 \}$ or $S^3 = \{ q \in \mathbb{H} : |q|^2=1\}$
\end{center}

where $|q|^2=a^2+b^2+c^2+d^2$ if $q=a \mathbf{1} +b \mathbf{i} + c\mathbf{j} + d \mathbf{k}$.

In other word, the sphere $S^3$ is a set of unit quaternions which has dimension $3$ by the condition $|q|^2=1$. (i.e., such a group is called $Sp(1)$.) Then there exists the isomorphism from the unit quaternions $S^3$ to $SU(2)$ and $SU(2)$ is topologically equivalent to 3-sphere $S^3$. (i.e., $SU(2) \simeq S^3$.)

Since $SO(3) \simeq D^3/ \sim$ and it is same as  all of $S^3$ with antipodal points identified, $SU(2) \simeq S^3$ is the universal covering space of $SO(3)$. (i.e., $n$-sphere with $n \geq 2$ is simply-connected.)

In fact, we can define the map $SU(2) \rightarrow SO(3)$ as follows.

Let $x$ be a pure quaternion which takes the form of  $x=b\textbf{i}+c\textbf{j}+d\textbf{k}$ and let $q \in SU(2)$. (i.e., $q$ is an unit quaternion.)

Then, the action or the linear map $\mathbb{R}^3 \rightarrow \mathbb{R}^3$ of $SU(2)$ defined by

\begin{center}
$x \rightarrow qxq^{-1}$ (or $q^{-1}xq$)
\end{center}
preserves the standard inner product and so lies in $SO(3)$. (i.e., $qxq^{-1}$ has the same norm as $x$ and rotations $SO(3)$ preserve the norm.) Since $q$ and $-q$ produces the same rotation, the kernel of the map is $\{\pm 1\}$ and its cosets are the sets $\{\pm q\}$. Therefore, $SO(3) \simeq SU(2) / \{\pm 1\}$. Then every element of $SO(3)$ corresponds to a pair of elements of $SU(2)$ which differ by sign and so $SU(2)$ is called the \emph{double cover} of $SO(3)$. 

As we saw earlier, every element of $SO(3)$ except the identity can be described by a rotation axis $\mathbf{n}$ and a rotation angle $\theta$ and two pairs $(\mathbf{n}, \theta)$ and $(-\mathbf{n}, -\theta)$ represent the same rotation. Then the choice of one of these pairs is called \emph{the choice of spin} and every elements of $SU(2)$ except $\pm1$ can be described as a rotation of $\mathbb{R}^3$ together with a choice of spin. [1]

Now, $SO(3)$ has a periodicity of $2\pi$, while $SU(2)$ has a periodicity of $4\pi$. Furthermore, if we parametrize the matrix $U$ in $SU(2)$  in terms of a rotation axis $\mathbf{n}$ and a rotation angle $\theta$, we have $U(0, \mathbf{n})=1$ and $U(2\pi, \mathbf{n})=-1$.

In physics, it is known that if we rotate a spin state throgh an angle $2\pi$, we find that states with half-integer spin obtain a minus sign, while integer spin states are rotated into what we started with. It is related to the fact that $SU(2)$ has integer and half-integer spin representations, however $SO(3)$ has only integer spin representations.

In higher dimensions, there is a compact, connected (simply-connected for $n \geq 4$) group, the spin group $Spin(n)$ which is the double cover of $SO(n)$. (i.e., there exists a surjective homomorphism, $\rho : Spin (n) \rightarrow SO(n)$, whose kernel is $\{-1,1\}$.) $Spin(n)$ can be constructed as a subgroup of the invertible elements in the \emph{Clifford algebra} $Cl_{n}$. (i.e., the action of $Spin(n)$ on $\mathbb{R}^{n}$ is given in terms of multiplication in an algebra $Cl_{n}$.) For $n=3$, we have $Spin(3)$ which is isomorphic to $SU(2)$. $Spin(3) \simeq Sp(1) \simeq SU(2)$ where $Sp(1)$ is a symplectic group.

A detailed discussion about $SO(3)$ and $SU(2)$ can be found in many places such as [1,12].

Now, we know $SU(2)$ is simply-connected space and so there is only one class of paths, however $SO(3)$ has two classes of paths as we have seen because of the identification of antipodal points. Shulman [23] performed path integral on both $SU(2)$ and $SO(3)$ compared two results. The calculation he did is briefly summarized in [6] as follows.

By the analogy of the rigid body, the hamiltonian $H$ of a free particle on $SO(3)$ and $SU(2)$ can be written as

\begin{center}
$H=-\frac{\hbar^2}{2I} \bigtriangledown ^2$
\end{center}

where $I$ has the physical dimension of a moment of inertia and the radius of the curvature $R$ is $2$.

Let the Euler angles denoted by $(\phi, \theta, \psi)$. The metric tensor for $SU(2)$ and $SO(3)$ is $g_{\phi\phi}=g_{\theta \theta}=g_{\psi\psi}=1$, $g_{\psi\phi}=g_{\phi\psi}=\cos \theta$.

Then the fundamental line element can be expressed as

\begin{center}
$(ds)^2=g_{ij}dE_{i}dE_{j}=(d\phi)^2+(d\theta)^2+(d\psi)^2+2\cos\theta d\phi d\psi$
\end{center}

and we have laplacian:

\begin{center}
$\bigtriangledown^2 = \frac{\partial^2}{\partial \theta^2} +\cot \theta \frac{\partial}{\partial \theta}+\frac{1}{\sin ^2 \theta} \left( \frac{\partial ^2}{\partial \psi ^2}+\frac{\partial ^2}{\partial \phi^2} -2 \cos \theta \frac{\partial^2}{\partial \psi \partial \phi} \right)$
\end{center}
 
The normalized eigenfunctions of this laplacian are [6]\\

$\Phi ^{j}_{mk}$ (on $SU(2)$) $=\left( \frac{2j+1}{16 \pi^2} \right)^{1/2} D^{j*}_{mk} (\theta,\psi,\phi)$ for $j=0,\frac{1}{2},1,\frac{3}{2}, \ldots$\\

$\Phi ^{j}_{mk}$ (on $SO(3)$) $=\left( \frac{2j+1}{8 \pi^2} \right)^{1/2} D^{j*}_{mk} (\theta,\psi,\phi)$ for $j=0,1,2,3, \ldots$\\

with eigenvalues $E_{jmk}=\frac{\hbar^2}{2I} j (j+1)$.\\

where labels $j,m,k$ are related to the eigenvalues of angular momentum $J^2, J_{z}, J_{\xi}$ respectively ($J_{\xi}=\hat{n}_{\xi} \cdot \mathbf{J}$ where $\hat{n}_{\xi}$ points along the figure axis) and $D$'s form a matrix representation of $SU(2)$.

Then the propagator from a point $a$ at time $t_{a}$ to a point $b$ at time $t_{b}$ on $SU(2)$ can be expressed by

\begin{center}
$K_{SU(2)}(b,t_{b};a,t_{a})=\displaystyle\sum_{j,m,k} \langle \Psi^{j}_{mk} (b)| \exp\left(-\frac{iE \bigtriangleup t}{\hbar}\right) |\Psi^{j}_{mk} (a) \rangle$

$=\displaystyle\sum_{j,m,k} \Psi_{mk}^{j*} (b) \exp \left(-\frac{i\hbar(t_{b}-t_{a})}{2I} j(j+1) \right) \Psi^{j}_{mk} (a)$

\end{center}

Schulman expressed this propagator as the sum of two terms related to integer and half-integer spin respectively [6]:

\begin{center}
$K_{SU(2)}=K_{SU(2)} ($integer $j) +K_{SU(2)}($half-integer $j)$
\end{center}

and also found its relation with two partial propagators $K^{I}_{SO(3)}$ and $K^{II}_{SO(3)}$  related to class I and class II paths on $SO(3)$  respectively:

\begin{center}
$2K_{SU(2)}($integer $j) = K^{II}_{SO(3)} -K^{I}_{SO(3)}$

$2K_{SU(2)}($half-integer $j)=K^{II}_{SO(3)}+K^{I}_{SO(3)}$

\end{center}

Note that class I and class II paths are called in the opposite way in [6].

Therefore, we can obtain the propagator of a integer-spin by subtracting the contribution of homotopy class of paths I from path II, while the propagator of a half-integer spin can be obtained by adding the contributions of both homotopy classes of paths. Furthermore, we have
\begin{center}
$K_{SU(2)}=K^{II}_{SO(3)}$
\end{center}
from the above equations. It can be related to the fact that if we unwrap the identification of $SO(3)$ to obtain $SU(2)$, then class I paths will disappear and only class II paths will remain. (i.e., class II loops represent a rotation through $4\pi$ which is same as those in $SU(2)$.)

 This is one interesting example which indicates the connection between path integral and homotopy which describes the topological structure of space where path integral is performed.

\section{Aharonov-Bohm effect}

\quad The Aharonov-Bohm effect is a quantum mechanical phenomenon which shows that a charged particle is affected by an electromagnetic field even they are confined to a region where both the magnetic field and electric field are zero. Its experiment setup consists of a source of uniform energy charged, spinless particles (i.e., usually electrons are used and their spin or statistics don't play any role in the discussion of the Aharonov-Bohm effect.), a screen with double-slit in it, a screen to record the interference pattern and an infinitely long solenoid. The solenoid is a magnetic flux (an electromagnet) enclosed in an iron tube and the electromagnetic field is zero outside of the tube since the iron tube absorbs the electromagnetic field created by the magnetic flux. (Figure 16)

\begin{figure}[htbp]
 \begin{center}
  \includegraphics[width=72mm]{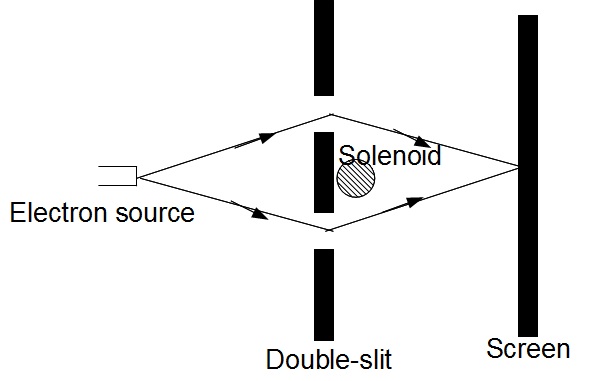}
 \end{center}
 \caption{Aharonov-Bohm effect}
 \label{fig:}
\end{figure}
\newpage
If we perform the experiment with no magnetic flux in the tube, then we have the standard interference pattern for the double-slit experiment. However, if we increase the magnetic flux, then there is a shift in the interference pattern despite the fact that electrons move through region where the electromagnetic field is zero.

It is known that we can understand the double-slit experiment or the Aharonov-Bohm effect using the concept of path integrals in multiply-connected spaces, which is discussed in some references such as [8,17]. Also, since there is a relation between path integral in multiply-connected spaces and homotopy as we discussed in the section 3, there are some attempts to study the Aharonov-Bohm effect using homotopy theory. [20,24] In this section, we will review these attempts.

Now, we can assume the configuration space of electrons $X$ is an annulus where an infinitely long solenoid perpendicular to the plane $\mathbb{R}^2$ in which electrons move around is located in the centre.

The theorem 3.1 for path integral on multiply-connected space is usually only valid for spinless particles. However, we treat electrons as spinless particles here as explained above. Since an annulus is contractible to a circle, the fundamental group $\pi_1 (X,x)=\pi_1 (S^1,x)$ is the additive group of the integers $\mathbb{Z}$. Therefore, using the theorem 3.1, the probability amplitude from the position of the source $\mathbf{r}_{i}$ to some position on the screen $\mathbf{r}_{f}$ for the Aharonov-Bohm effect is given by

\begin{equation}
K(\mathbf{r}_{f} , t_{f} ; \mathbf{r}_{i} , t_{i}) = \displaystyle\sum_{n=-\infty}^{\infty} e^{-in\delta} K_{n} (\mathbf{r}_{f},t_{f} ; \mathbf{r}_{i} ,t_{i})
\end{equation}
with $\delta \in [0, 2\pi)$ and $n \in \mathbb{Z}$. $K_{n}$ are partial probability amplitudes obtained by integrating over paths in same homotopy class. By using the same notation given in [24], $n$th homotopy class contains paths which wind around the solenoid $n-1$ times counterclockwise for $n \geq 2$, while wind $|n|$ times clockwise for $n \leq -1$. Also, we define the 0th homotopy class ($n=0$) and 1st homotopy class ($n=1$) contains paths which do not wind around the solenoid but stay above and below it respectively. (Figure 17)

We can argue $\delta=0$ as follows. [20] Let $\mathbf{r}_{f}$ and $\mathbf{r}'_{f}$ be two points on the screen which are symmetrically located with respect to the axis connecting the source $\mathbf{r}_{i}$ to the centre of the solenoid and the centre of the screen $\mathbf{r}^0_{f}$. (Figure 18) Let $[q_{n} (\mathbf{r}_{i} , \mathbf{r}_{f})]$ be the $n$th homotopy class of paths from $\mathbf{r}_{i}$ to $\mathbf{r}_{f}$. Now, let $q$ be a path in $[q_{n} (\mathbf{r}_{i} , \mathbf{r}_{f})]$ and define $q'$ which is a path obtained by reflecting $q$ about the axis we introduced above. (Figure 18)
\newpage
\begin{figure}[htbp]
 \begin{center}
  \includegraphics[width=90mm]{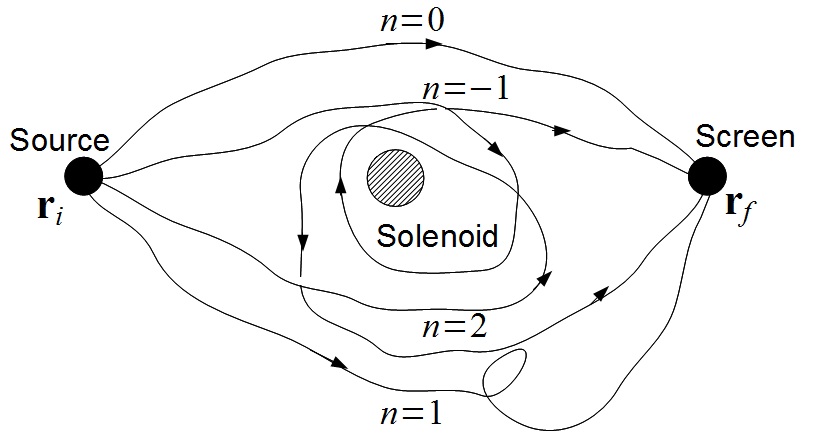}
 \end{center}
 \caption{$n$th homotopy classes of paths}
 \label{fig:}
\end{figure}
\begin{figure}[htbp]
 \begin{center}
  \includegraphics[width=70mm]{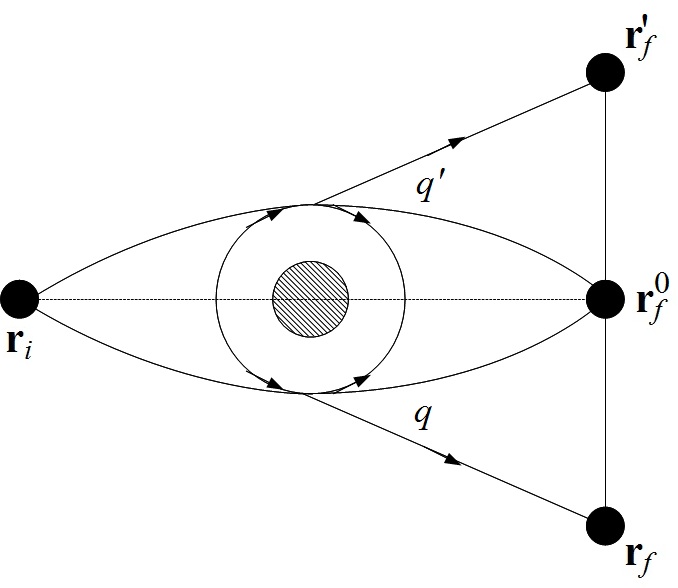}
 \end{center}
 \caption{$q$ and $q'$}
 \label{fig:}
\end{figure}
Then we have

\begin{center}
$q' \in [q_{-n+1} (\mathbf{r}_{i} , \mathbf{r}'_{f})]$
\end{center}

Let us consider the case in which the magnetic flux in the tube is zero. Then, electrons are free particles except that there is a solenoid in their configuration space. In this case, the action integral along $q$ and $q'$ is same since there is no external fields and so we have 

\begin{center}
$K_{n} (\mathbf{r}_{f},t_{f} ; \mathbf{r}_{i} ,t_{i}) = K_{-n+1} (\mathbf{r}'_{f}, t_{f} ; \mathbf{r}_{i} , t_{i})$
\end{center}

Therefore

\begin{center}
$K(\mathbf{r}_{f}, t_{f} ; \mathbf{r}_{i}, t_{i}) =\displaystyle\sum_{n=-\infty}^{\infty} \exp (-in\delta) K_{n} (\mathbf{r}_{f}, t_{f} ; \mathbf{r}_{i} , t_{i})$

$K(\mathbf{r}'_{f},t_{f} ; \mathbf{r}_{i} , t_{i}) = \displaystyle\sum_{n=-\infty}^{\infty} \exp [(-i\delta (-n+1)] K_{n} (\mathbf{r}'_{f}, t_{f} ; \mathbf{r}_{i} , t_{i})$
\end{center}

If $\mathbf{r}_{f}=\mathbf{r}'_{f}=\mathbf{r}^0_{f}$, we can sum over the contributions of the paths passing over the solenoid (first terms) and passing under the solenoid (second term): \\

$K(\mathbf{r}^0_{f} ,t_{f} ; \mathbf{r}_{i} , t_{i}) $
\begin{center}
$=\displaystyle\sum_{n=1}^{\infty} \exp [-i\delta (-n+1)]K_{-n+1}(\mathbf{r}^0_{f} ,t_{f} ; \mathbf{r}_{i} , t_{i}) +\displaystyle\sum_{n=1}^{\infty} \exp(-i n\delta)K_{n}(\mathbf{r}^0_{f} ,t_{f} ; \mathbf{r}_{i} , t_{i})$

$=\displaystyle\sum_{n=1}^{\infty} \exp [-i\delta (-n+1)] K_{n} (\mathbf{r}^0_{f} ,t_{f} ; \mathbf{r}_{i} , t_{i})+\displaystyle\sum_{n=1}^{\infty} \exp(-in\delta)K_{n}(\mathbf{r}^0_{f} ,t_{f} ; \mathbf{r}_{i} , t_{i})$

$=\displaystyle\sum_{n=1}^{\infty} \exp [-i\delta(-n+1)]\times [1+\exp (-i\delta(2n-1))]K_{n} (\mathbf{r}^0_{f} ,t_{f} ; \mathbf{r}_{i} , t_{i}) $
\end{center}

As it was mentioned earlier, we have the standard interference pattern for the double-slit experiment without magnetic flux. Then, it is known that the interference pattern has the following properties.\\

(a) The interference pattern is symmetric (i.e., the bright (constructive) and dark (destructive) interferences are symmetric about the centre of the screen $\mathbf{r}^0_{f}$.) :

\begin{center}
$|K(\mathbf{r}_{f}, t_{f} ; \mathbf{r}_{i}, t_{i})|^2 = |K(\mathbf{r}'_{f},t_{f} ; \mathbf{r}_{i} , t_{i})|^2$
\end{center}

(b) At the centre of the screen, we have the bright or dark interference according to the distance between solenoid or slits and the screen. The intensity is proportional to $|K(\mathbf{r}^0_{f}, t_{f} ; \mathbf{r}_{i}, t_{i})|^2$.\\

(a) implies that $\delta=0$ or $\delta=\pi$. If we substitute them into $K(\mathbf{r}^0_{f} ,t_{f} ; \mathbf{r}_{i} , t_{i})$, we have

\[
K(\mathbf{r}^0_{f} ,t_{f} ; \mathbf{r}_{i} , t_{i})=
\begin{cases}
2\displaystyle\sum_{n=1}^{\infty}K_{n}(\mathbf{r}^0_{f} ,t_{f} ; \mathbf{r}_{i} , t_{i}), &\text{if $\delta=0$}\\
0, &\text{if $\delta=\pi$}
\end{cases}
\]

Then, we need to choose $\delta=0$ from (b). Therefore, we use the trivial representation of the fundamental group and substitute the result into (5.1) to obtain

\begin{equation}
K(\mathbf{r}_{f},t_{f};\mathbf{r}_{i},t_{i})=\displaystyle\sum_{n=-\infty}^{\infty} K_{n}(\mathbf{r}_{f}, t_{f} ; \mathbf{r}_{i},t_{i})
\end{equation}

Now, we study how the electromagnetic field in the solenoid plays a role in the Aharonov-Bohm effect. The reference for the following discussion is [22]. Although the electric field $\mathbf{E}$ and the magnetic field $\mathbf{B}$ are zero outside the solenoid, the magnetic vector potential $\mathbf{A} \not= 0$ and this causes the effect. In cylindrical polar coordinates, $\mathbf{A}$ which gives a solenoidal magnetic field has the following form.\\

(i) Inside the solenoid
\begin{center}
$A_{r}=A_{z}=0, A_{\phi} =\frac{Br}{2}$
\end{center} 

(ii) Outside the solenoid

\begin{center}
$A_{r}=A_{z}=0, A_{\phi}=\frac{BR^2}{2r}$
\end{center}
where $R$ is a radius of the solenoid and the magnetic field $\mathbf{B}$ is given by\\

(iii) Inside the solenoid
\begin{center}
$B_{r}=B_{\phi}=0, B_{z}=B$
\end{center}

(iv) Outside the solenoid
\begin{center}
$\mathbf{B}=0$
\end{center}

Since
\begin{center}
$\bigtriangledown \times \mathbf{A} =\mathbf{B}=0$
\end{center}

there exists a scalar function $\chi$ such that $\mathbf{A}=\bigtriangledown \chi$ in a small region by Poincar\'{e} lemma.

From (ii), we have
\begin{center}
$A_{\phi} =\frac{1}{r} \frac{\partial \chi}{\partial \phi} =\frac{BR^2}{2r}$
\end{center}
and obtain $\chi = \frac{BR^2}{2} \phi$. Since $\chi$ increases by $\pi R^2 B$ when $\phi \rightarrow \phi +2\pi$, it is not a single-valued (many-valued) function. $\chi$ is a many-valued function since it is defined on a multiply-connected space and if $\chi$ were single-valued, then $\mathbf{B}=\bigtriangledown \times \mathbf{A} = \bigtriangledown \times (\bigtriangledown \chi)=0$ everywhere and we can not have any magnetic flux. Therefore, the Aharonov-Bohm effect occurs only if the configuration space is multiply-connected.

Now, the Maxwell action of electromagnetism is represented in terms of the electromagnetic tensor:

\begin{center}
$S_{EM}=-\frac{1}{4} \int F^{\mu \nu} F_{\mu \nu} d^4 x$
\end{center}

where $F_{\mu\nu}=\partial_{\mu} A_{\nu} - \partial _{\nu} A_{\mu}$.

Then the gauge transformation

\begin{center}
$A_{\mu} \rightarrow A_{\mu}  -\partial _{\mu} \chi$
\end{center}
leaves the action unchanged. Therefore, $A_{\mu}=\partial_{\mu} \chi$ is actually a gauge transform of the vacuum $A_{\mu}=0$.

Those transformations form a gauge group known as the unitary group $U(1)$. (i.e., A gauge transformation for a group is written by $A_{\mu} = UA_{\mu}U^{\dagger}-iU\partial_{\mu}U^{\dagger}$ which gives the above transformation if $U=e^{i\chi(x)} \in U(1)$.) 

It is known that $U(1)$ is isomorphic to a circle $S^1 =\{z \in \mathbb{C}| |z|=1 \}$. Thus, $\pi_1 (U(1),x)=\pi_1 (S^1, x) =\mathbb{Z}$. Now, the gauge function $\chi$ can be viewed as a mapping from the group space $G$ onto configuration space $X$:

\begin{center}
$\chi : G \rightarrow X$
\end{center}
In this case, $G=U(1) \simeq S^1$ and $X \simeq S^1$ and so we have
\begin{center}
$\chi : S^1 \rightarrow S^1$
\end{center}
Actually, we can define the homotopy groups using the above method as follows.
Let $[Y,X]$ be the set of all homotopy classes of continuous maps from $Y$ onto $X$. Then we have

\begin{center}
$\pi_1 (X,x) =[S^1, X]$
\end{center}
is the fundamental group (first homotopy group) of $X$. Furthermore,

\begin{center}
$\pi_{n} (X, x) = [S^{n} , X]$
\end{center}
is the $n$th homotopy group of $X$ and it is an Abelian group for $n \geq 2$.

Since the gauge functions $\chi$ are maps from $S^1$ onto $S^1$, we have

\begin{center}
$[S^1 , S^1] = \pi_1 (S^1) =\mathbb{Z}$
\end{center}
Therefore, nonzero vector potential $A_{\mu}$ which causes the Aharonov-Bohm effect is derived from a gauge function $\chi$ which maps the gauge space onto the configuration space. [22]

For example, if the gauge group were a simply-connected space such as $SU(2)$ or $SU(3)$, then we have
\begin{center}
$\pi_1 (SU(2))=\pi_1 (SU(3)) =1$
\end{center}
and a gauge function $\chi$ is constant as a result. This gives $A_{\mu}=0$ and so we don't have Aharonov-Bohm effect. In conclusion, the Aharonov-Bohm effect exists since the gauge group $U(1)$ and the configuration space are multiply-connected.

Now, we return to the discussion about the probability amplitude. As we saw in the previous section, it is often convenient to peform the path integral in the universal covering space (simply-connected) of the configuration space $X$ (multiply-connected) and sum over the contributions from different homotopy classes of paths in $X$. The universal covering space $\tilde{X}$ of $X$ is essentially the same as the Riemann surface for the logarithm [24] where for each $n$, we have the pre-images $\mathbf{r}_{f}^{(n)}=(r_{f}, \theta +2n\pi) \in \tilde{X}$ of $\mathbf{r}_{f}=(r_{f}, \theta _{f}) \in X$ ($0\leq \theta_{f} <2\pi$) which is the point on the screen.

The Lagrangian of an electron moving in electromagnetic field is given by

\begin{center}
$\mathcal{L} (\mathbf{r}_{e} , \dot{\mathbf{r}}_{e} , t) = \frac{1}{2} m \dot{\mathbf{r}}_{e}^2 + \frac{e}{c} \mathbf{A} \cdot \dot{\mathbf{r}}_{e}$
\end{center}
where $\mathbf{r}_{e}$ and $\dot{\mathbf{r}}_{e}$ is a position and velocity of an electron respectively, $e$ is an electron charge and $c$ is a speed of light.

Since $\tilde{X}$ is simply-connected, we have a scalar function $\chi$ such that globally $\mathbf{A} (\mathbf{r}) =\bigtriangledown \chi (\mathbf{r})$ by Poincar\'{e} lemma. Therefore, the probability amplitude $\tilde{K}$ in $\tilde{X}$ can be written as

\begin{equation}
\tilde{K}(\mathbf{r}_{f},t_{f};\mathbf{r}_{i},t_{i})=\displaystyle\sum_{n=-\infty}^{\infty} \exp \left(i\frac{e}{\hbar c}[\chi(\mathbf{r}_{f}^{(n)})-\chi( \mathbf{r}_{i})]\right) K_{n}(\mathbf{r}_{f}, t_{f} ; \mathbf{r}_{i},t_{i})
\end{equation}

Since $\chi=\frac{BR^2}{2} \phi$, the probability amplitude $K_{AB}$ for the Aharonov-Bohm effect is

\begin{center}
$K_{AB}(\mathbf{r}_{f},t_{f};\mathbf{r}_{i},t_{i})=\displaystyle\sum_{n=-\infty}^{\infty} \exp \left(i\frac{eBR^2}{2\hbar c}[\phi_{f}-\phi_{i}+2n\pi]\right) K_{n}(\mathbf{r}_{f}, t_{f} ; \mathbf{r}_{i},t_{i})$
\end{center}
\begin{equation}
=\displaystyle\sum_{n=-\infty}^{\infty} \exp \left(i\frac{e\Phi}{2\pi \hbar c}[\phi_{f}-\phi_{i}+2n\pi]\right) K_{n}(\mathbf{r}_{f}, t_{f} ; \mathbf{r}_{i},t_{i})
\end{equation}
where $\Phi$ is the magnetic flux of the solenoid and $B=\frac{\Phi}{\pi R^2}$.

Thus, the interference pattern changes periodically as we change the magnetic flux $\Phi$. [20] gives the further calculation for $K_{n}$. 

\section{Summary}

\quad In this review, we discussed the application of homotopy theory in path integrals found in studies by Schulman, Laidlaw, DeWitt, Morandi and Menossi. There are many applications of algebraic topology to physics. For example, Dirac monopoles can be studied in the context of path integral on the multiply-connected space. [17] Also, there are many reviews on the studies of solitons and monopoles using mathematical approach such as algebraic topology or geometry. [2,18]

\section*{Acknowledgements} 
\quad I wish to thank John C. Wood, who taught me homotopy theory in a course at the University of Leeds and gave me helpful advices on this review and Richard MacKenzie, who taught me about physics involving path integrals on the multiply-connected spaces. I also would like to thank P. C. E. Stamp for giving me the opportunity of talking about this review.

\end{document}